# OPERATIVE AND PROCEDURAL COOPERATIVE TRAINING IN MARINE PORTS

Francesco Longo[a], Alessandro Chiurco[b], Roberto Musmanno[c], Letizia Nicoletti[d],

[a][b][c][d] DIMEG, University of Calabria, Rende (CS), Italy

f.longo@unical.it[a], a.chiurco@unical.it[b], musmanno@unical.it[c], letizia.nicoletti@unical.it[d]

**ABSTRACT**
This article faces the problem of operative and procedural cooperative training in marine ports with particular attention to harbour pilots and port traffic controller. The design and development of an advanced system, equipped with dedicated hardware in the loop, for cooperative training of operators involved in the last mile of navigation is presented. Indeed, the article describes the software and hardware development of a distributed and interoperable system composed by two simulators (the bridge ship simulator and control tower simulator). Multiple problems are faced and solved including (i) the motion of the ship at sea that is based on a 6 Degree Of Freedom (DOF) model for surge, sway and yaw and closed form expressions for pitch, roll and heave and its validation; (ii) the development of the 3D geometric models and related virtual environments of a real marine port and vessel (to provide the trainees with the sensation to experience a real port and ship environment); (iii) the design of a bridge ship replica, the bridge hardware integration and the design of the visualization system; (iv) the design and development of the control tower simulator; (v) the integration of the bridge ship simulator and control tower simulator through the IEEE 1516 High Level Architecture standard for distributed simulation.

## 1. INTRODUCTION

Large ships manoeuvres in the aim of the entry to or exit from the harbour area can be complex and dangerous operations for several reasons. Indeed, in standard traffic conditions, harbour pilots and port traffic controllers deal with a great number of vessels of different sizes (i.e. small motorboats, huge container carriers, cruise ships etc.). In addition, manoeuvres times (and reaction times) of vessels are very slow compared to other vehicles especially for large vessels. Indeed, due to their prominent size and mass, in case of mistakes and accidents, large vessels may cause enormous damages (as clearly demonstrated by the recent accident on May 7th, 2013 of the Italian containership Jolly Nero in the port of Genoa, Italy, check http://www.bbc.com/news/world-europe-22444421 for further information).

In large facilities like marine ports like in many other industrial facilities, operative and procedural training has widely proved to be an indispensable approach that can be effectively supported by simulation based systems (Bruzzone et al., 2010; Longo, 2011; Merkuryev and Bikovska, 2012); as far as the ship manoeuvres in the port area are concerned, simulation based systems can be profitably used for ship pilots training purposes, for procedures definition, evaluation and testing, for understanding vessels interactions within the port area, for evaluating the effects on ships of adverse weather conditions (including wind, sea waves and marine currents). As a matter of facts, simulation based system allows a greater standardization and effectiveness of training procedures. The trainees can control virtual ships in any critical or dangerous condition; they can perform both standard and non-standard manoeuvres and therefore they are trained to handle those situations that cannot be recreated during traditional training sessions in a real ship.

Therefore simulation based training is also characterized by a relevant reduction of direct costs. These costs reduction is related to the increase of productivity when workers (in this case ship pilots) are well trained to perform their job. There is also a reduction of indirect costs caused by accidents during training and normal activities as well as the possibility of collecting a large amount of data about trainees' performance in a controlled simulated environment. Moreover, simulation-based training is fruitful not only for beginners but also for expert operators. For instance, it can be useful to have the first approach with a new ship or to learn how to put in practice the procedures currently adopted in a specific port. Moreover, the simulators can be used to understand how performing ship manoeuvres in critical conditions (i.e. strong wind and/or marine currents), how carrying out complex manoeuvres with the help of tugboats, how performing mooring operations and control the ship when close to the assigned berth area. A key feature of the simulation approach is the greater level of immersion it can provide compared to traditional training approaches. Indeed, immersion makes the training experience more captivating and therefore contributes to maximize the amount of information the trainees are able to acquire, the skills that can be developed and mostly the ability to transfer lesson learned to the real system. Basically, immersion can be diegetic and situated or intra-diegetic. The former occurs when the player gets absorbed into the virtual experience while the latter goes beyond the diegetic immersion and implies the player total engagement that is the illusion of existing and acting within the virtual environment. Needless to say that, in both cases, the virtual environment where the trainee acts and the hardware tools used to interact with the environment itself play a crucial role. To this end, the research proposed in this article has been focused on the development of a fully operational prototype system for harbour pilots and port traffic controllers training including not only the distributed virtual environments, but also a suitable hardware tools design and integration. The training system presented in this article is currently installed in the MSC-LES lab (Modeling and Simulation Center-Laboratory of Enterprise Solutions) at the University of Calabria.

Indeed, the literature review confirms that there is quite a large amount of research works that show how simulation has already been successfully applied to support decision making (Bruzzone et al. 2012-a) and training of operators working within the port area, i.e. cranes, trucks, straddle carriers operators, etc. (Kim, 2005). Many simulators are intended to quay cranes operators' training (Wilson et al., 1998; Huang, 2003; Daqaq, 2003; Rouvinen et al., 2005) and specific research works have also investigated the training for supporting security procedures integration in the marine port operations (Longo and Bruzzone 2005, Longo, 2010 and Longo 2012). To this end, it is worth mentioning that a comprehensive survey of research projects dealing with advanced simulation systems for operators training in marine ports is the main deliverable of the OPTIMUS project (Operational Port Training Models Using Simulators, financed by the European Community, OPTIMUS project 2009). The OPTIMUS project provides a detailed description of many commercial simulators for marine port operators training. Such simulators include crane simulators mainly.

In addition, innovative approaches based on interoperable simulation have been proposed. Specific examples regards the training of marine ports operators, e.g. Bruzzone et al. 2008, Bruzzone et al. (2011-a), Bruzzone et al. (2011-b) propose interoperable simulators (based on the High Level Architecture, HLA, standard) for different container handling equipment (gantry cranes, transtainers, reach stackers, trucks, etc.) offering advanced solutions in terms of external hardware, e.g. motion platforms, different types of external controllers from joysticks to wheels and pedals and even containerized solutions.

As for ship simulators, interesting applications can be found in Ueng et al. (2008), Sandaruwan et al. (2009), Sandaruwan et al. (2010) and Yeo et al. (2012). Although in such works an immersive visualization system and the dynamic behaviour of the ship have been implemented, there is still substantial room for improvement above all in terms of ship motions predictions and validation. Moreover, in these works no attention has been paid on the last/second last mile of navigation (including manoeuvring and mooring operations in the port area) that, as explained before, it is as important as offshore navigation. A review of the state of the art related to traffic controllers and ships pilots training in marine ports can be found in Bruzzone et al. (2012-b).

Indeed, the main goal of this paper is to present an advanced simulation based system, equipped with dedicated hardware in the loop, for training, safety and security of operators involved in the last mile of navigation. The system includes a ship bridge simulator (with a full replica of a ship bridge), a full control tower simulator and a tugboat bridge simulator (with a full replica of a tugboat bridge; the last simulator is not described as part of this article). The proposed training system is conceived in order to provide its users with a realistic experience thanks to the possibility of experiencing a joint and cooperative training environment. To this end, the three simulators have been integrated according to the High Level Architecture standard (IEEE HLA 1516). As a result, pilots can exercise their operational and manoeuvring skills, became acquainted with the behaviour of the ship and with the effects of their interaction patterns. Moreover, the proposed system allows the harbour pilots to learn the procedures that are currently adopted in a specific port, and decision makers to design and test new procedures.

Along the article different problems are presented and solved, namely: (i) the sea-keeping problem in terms of implementation and validation of the ship motion equations at sea for a 6 Degrees of Freedom (DOF) model; (ii) the design and development of all the 3D geometric models and virtual environments; (iii) the design of a bridge ship replica, the bridge hardware integration and the design of the visualization system; (iv) the design and development of the control tower simulator; (v) the integration of the bridge ship simulator and control tower simulator through the IEEE 1516 High Level Architecture standard for distributed simulation.

The prototype system presented in this article has been developed within the framework of the on-going research project: HABITAT (Harbour Traffic Optimization System). This project is co-founded by the Italian Ministry of Education, University and Research as part of the PON01_01936 research project.

Before going into details, the paper is organized as follows: section 2 discusses the sea keeping problem and explains the 6 DOF model used to simulate the ship motion at sea. section 3 presents the 3D geometric models and the virtual environments based on the Port of Salerno (Italy). Section 4 proposes the design of the ship bridge replica, the hardware integration and the visualization system, while section 5 presents the control tower simulator. Finally section 6 show the overall system prototype including the ship bridge simulator and the control tower simulator and the last section summarizes the scientific contribution of the paper and points out some aspects for future development.

## 2. THE SEA-KEEPING PROBLEM: THE SHIP MOTION EQUATIONS

The ship included as part of the ship bridge simulator is a commercial containership based on the Kriso containership model, whose main particulars are summarized as follows:

- Hull
    - Length between perpendiculars 230.0 m
    - Length water line 232.5 m
    - Breath 32.2 m
    - Depth 19.0 m

- Displacement 52030 m³
- Coefficient block 0.651
- Rudder
  - Type semi-balanced horn rudder
  - Surface of rudder 115 m²
  - Lateral area 54.45 m²
  - Turn rate 2.32 deg/s
- Propeller
  - Number of blades 5
  - Diameter 7.9 m
  - Pitch ratio, P/D (0.7R) 0.997
  - Rotation Right hand

A 6 DOF mathematical model is used to reproduce the ship motion at sea. In particular, for surge, sway and yaw, the Manoeuvring Mathematical Modelling Group model (MMG), NMG 1985, has been used (and tuned). Such model takes the name from the Japanese research group that implemented it for the first time between 1976 and 1980. Hence, the MMG group (1985) defined for the first time a prediction method for ship manoeuvrability. Afterwards, Kijima and Nakiri (1999), taking into account the effects of the stern, proposed the approximate formulas for evaluating the hydrodynamic forces acting during manoeuvring motions. As for the hydrodynamic coefficients, empirical formulas are given in Rhee and Kim (1999) and Lee et al. (2003). Moreover, Yoshimura (1986) and Perez et al. (2006) describe how some parameters and dimensions influence manoeuvrability characteristics while Hasegawa et al. (2006) have discussed the course-keeping in windy condition. These findings have been integrated within the MMG model that includes three equations of motion (1, 2, 3) for surge, sway and yaw respectively based on the Newton's second law and according to the reference system depicted in figure 1.

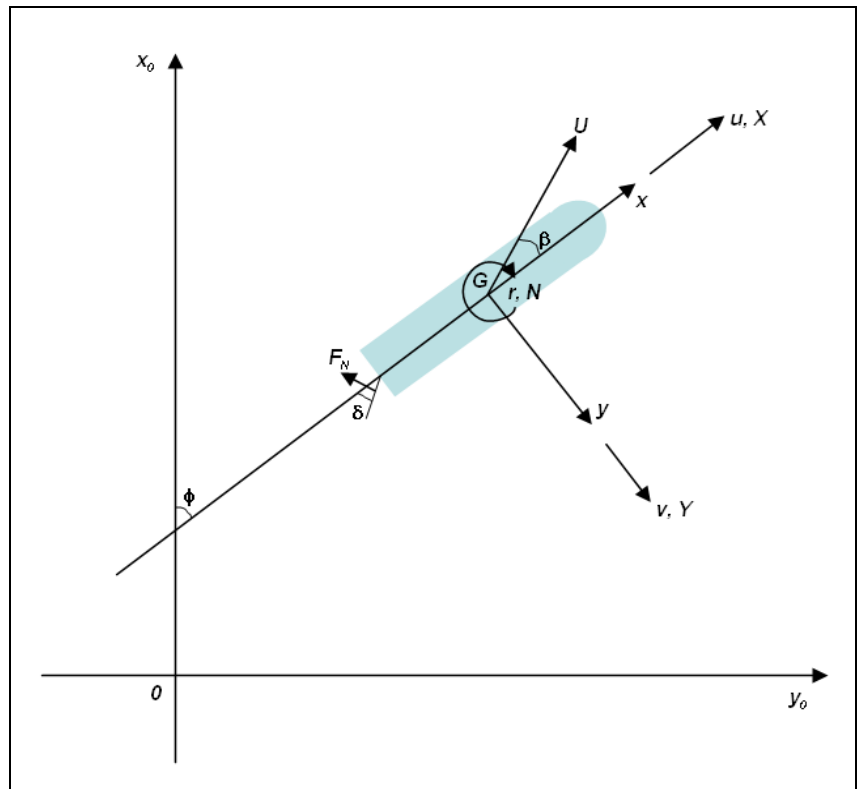

Figure 1 – Reference system for ship motion equations (MMG, 1985)

$$(m + m_x)\dot{u} - mvr = X \tag{1}$$

$$(m + m_y)\dot{v} + mur = Y \tag{2}$$

$$(I_{zz} + i_{zz})\dot{r} = N - x_G Y \tag{3}$$

In figure 1 and equations 1, 2 and 3:

- $m$ is the mass of the ship;
- $m_x$ and $m_y$ are the added mass in x and y direction respectively;
- $I_{zz}$ is the moment of inertia;
- $i_{zz}$ is the added moment of inertia around z;
- $u$ is the surge speed;
- $v$ is the sway speed;
- $U$ is the ship total speed
- $r$ is the rate of turn;
- $G$ is the center of gravity of the ship
- $x_G$ is the distance from amidship to the centre of gravity of the ship
- $X$ and $Y$ are respectively the total external surge and sway forces;
- $N$ is the yaw moment.
- $\delta$ is the rudder angle
- $\phi$ is the angle between the ship heading and the vertical axis of the reference system
- $F_N$ is the normal force applied to the rudder
- $\beta$ is the angle between the ship total speed and the ship heading

Added masses and added moment of inertia can be found using the equations proposed by Hooft and Pieffer (1988). In some equations non-dimensional variables are used, they are marked with the prime symbol. The external forces are those produced by the hull, the propeller and the rudder; they are marked with the subscripts H, P and R respectively, therefore surge sway and yaw can be also expressed as shown in equations 4, 5 and 6.

$$X = X_H + X_P + X_R \tag{4}$$

$$Y = Y_H + Y_R \tag{5}$$

$$N = N_H + N_R \tag{6}$$

Since the hull of the ship has a complex shape it is quite difficult to calculate hull forces. To this end, the equations 7, 8 and 9 (proposed in the Proceedings of the 23rd ITTC 2002) have been used.

$$X'_H = -\left(X'_0 + \left(X'_{vr} - m'_y\right)v'r'\right) \tag{7}$$

$$Y'_H = Y'_v v' + (Y'_r + m'_x)r' + Y'_{vvv} v'^3 + Y'_{vvr} v'^2 r' + Y'_{vrr} v' r'^2 + Y'_{rrr} r'^3 \tag{8}$$

$$N'_H = N'_v v' + (N'_r + m'_x)r' + N'_{vvv} v'^3 + N'_{vvr} v'^2 r' + N'_{vrr} v' r'^2 + N'_{rrr} r'^3 \tag{9}$$

It is worth noticing that the higher order terms have been omitted in the surge force equation because their influence is negligible and an appropriate formula to calculate the coefficients is not available yet. Equations 7, 8 and 9 define a relation between velocities and hull forces with hydrodynamic non-dimensional coefficients. Such coefficients are normally obtained through empirical tests, however a set of semi-empirical equations can be found in Lee et al. (2003). Therefore, the total non-dimensional resistance, $X'_0$, has been calculated as shown in equation 10 where $C_T$ is the total resistance coefficient (obtained from model resistance tests), Holtrop and Mennen, 1982, $S$ is the wetted surface, $L$ is the length between the perpendiculars, and $d$ is the draft.

$$X'_0 = \frac{C_T S}{Ld} \tag{10}$$

On the other hand, the equations for non-dimensional variables are given in 11, 12, 13, 14 and 15 according to Kijima (1993).

$$m', m'_x, m'_y = m, m_x, \frac{m_y}{0.5\rho L^2 d} \tag{11}$$

$$X', Y' = X, \frac{Y}{0.5\rho L d U^2} \tag{12}$$

$$N' = \frac{N}{0.5\rho L^2 d U^2} \tag{13}$$

$$r' = \frac{rL}{U} \tag{14}$$

$$v' = \frac{v}{U} \tag{15}$$

In equations 11, 12, 13, 14 and 15, $\rho$ is water density and $U$ is the ship speed that can be calculated according to 16.

$$U = \sqrt{u^2 + v^2} \tag{16}$$

The force generated by the propeller is obtained using the equation 17 from Kijima(1993) where $n$ is the propeller rate expressed in RPM, $t$ is the suction coefficient, $D_p$ is the propeller diameter and $K_T$ is the propeller thrust coefficient.

$$X_P = (1-t)\rho n^2 D_p^4 K_T \tag{17}$$

It is possible to express $K_T$ as a function of the propeller advance coefficient $J$, as shown in equations 18 and 19.

$$J = \frac{(1-w_p)u}{nD_p} \tag{18}$$

$$K_T = C1 + C2 + C3 J^2 \tag{19}$$

In equation 18, $w_p$ is the wake fraction, while in order to identify $C1$, $C2$ and $C3$ in equation 19 it is necessary to use the least squares method on Wageningen B systematic series for the appropriate propeller. Lastly, the rudder forces are calculated according to equations 20, 21 and 22 taken from Kijima (1993).

$$X'_R = -(1-t_R)F'_N \sin\delta \tag{20}$$

$$Y'_R = -(1-a_H)F'_N \cos\delta \tag{21}$$

$$N'_R = -(x'_R - a_H x'_H)F'_N \cos\delta \tag{22}$$

where:
- $t_R$ is the rudder drag coefficient;
- $F'_N$ is the normal force applied to the rudder;
- $a_h$ is a coefficient that expresses the interaction between rudder and hull forces;
- $x'_R$ is the non-dimensional coordinate of the centre of lateral force along the x-axes;
- $x'_H$ is the non-dimensional coordinate of the centre of additional lateral force along the x-axis (such values can be evaluated by using the equations given in Kijima and Tanaka, 1993);
- $\delta$ is the rudder angle.

As far as the three remaining degree of freedom are concerned (heave, pitch and roll), they have been modelled according to Jensen, (2001) and Jensen et al (2004). The proposed model is based on simplified equations, one for each DOF, devoted to work out ship motions in regular waves. Basically, in these equations, the coupling terms are neglected and the sectional added mass is equal to the displaced water.

$$2\frac{kT}{\omega^2}\ddot{\omega} + \frac{A^2}{kB\alpha^3\omega}\dot{\omega} + \omega = aF\cos(\bar{\omega}t) \qquad (23)$$

$$2\frac{kT}{\omega^2}\ddot{\theta} + \frac{A^2}{kB\alpha^3\omega}\dot{\theta} + \theta = aG\sin(\bar{\omega}t) \qquad (24)$$

$$\left(\frac{T_N}{2\pi}\right)^2 C_{44}\ddot{\varphi} + B_{44}\dot{\varphi} + C_{44}\varphi = Ma\cos(\bar{\omega}t) \qquad (25)$$

The equations 23 and 24 are related to heave and pitch respectively, while equation 25 refers to roll motions. Here, derivation with respect to time is denoted by a dot, $k$ is the wave number, $\omega$ is the wave frequency, $B$ is the ship breadth, $T$ is the ship draught, $\bar{\omega}$ is the frequency of encounter, $a$ is the wave amplitude and $A$ is the sectional hydrodynamic damping that can be evaluated according to Yamamoto et al. (1986). $F$ and $G$ are the forcing functions whose values can be worked out according to Jensen et al. (2004). As for roll, $\varphi$ is the roll angle, $T_N$ is the natural period for roll, $B_{44}$ is the ship hydrodynamic damping, $C_{44}$ is the restoring moment coefficient and $M$ is the roll excitation moment. The hydrodynamic damping coefficient $B_{44}$ can be found by applying the method described in Jensen et al. (2004). The roll excitation moment $M$ can be derived from the Haskind relation, while the restoring moment coefficient $C_{44}$ can be expressed as a linear function of the displacement $\Delta$, the transverse metacentric height $GM_T$ and the acceleration of gravity $g$ according to equation 26.

$$C_{44} = gGM_T\Delta \qquad (26)$$

It is worth noticing that this research work proposes a new application of the Jensen's et al. (2004) model. As a matter of facts, ship responses are calculated in the time domain instead of in the frequency domain as it was in the original formulation. Thus, this model has been used to simulate real-time the behaviour of the ships and to provide the simulation and visualization system with real-time data about heave, pitch and roll motions.

**2.1 Validation of the ship motion equations**

The above-discussed mathematical models have been extensively validated before being implemented within the ship simulator. As explained later on, the ship simulator provides the user with a real-time simulation in a 3D Virtual Environment; the validation of the ship motion equations in a such environment is time expensive, each single test would require too much time to be executed (e.g. a simulated circle test executed real-time would require different minutes as happens in a real circle test). To this end, ship motion equations validation has been carried out by using an ad-hoc developed tool (implemented by using Visual Studio 2008 and C++ programming language) equipped with a 2D animation and able to run the simulation fast-time for validation purposes.

This tool allows setting the most important parameters such as the number of iterations to be executed, the time between two iterations, the propeller rate, the initial speed and the rudder angle. In this way, it is possible to execute circle tests, zigzag tests and stop tests according to different configurations as shown in figure 2, 3, 4 and 5 and easily carrying equations validation according to experimental data and/or subject matter experts estimations.

During the execution of such tests, a plot of the ship motion evolution is drawn on the screen and sensible data as speed, drift angle, yaw rate, acceleration, heading, position are recorded on text files. This tool has proved itself as a very useful solution for ship motion equations validation. Since the preliminary tests carried out at the earlier validation stages were not in agreement with the empirical data that have been used as test-bed, the mathematical model has been carefully reviewed and some parameters, such as the dimensionless derivatives, where suitably calibrated based on tentatively approach. Even during calibration, the C++ tool has been used to assess whether the applied adjustments had improved the output results to achieve closer empirical evidence.

Figure 2 and 3 show the trajectory for two circle tests executed by using the validation tool, moreover they provide a comparison with empirical data (green squares) given by two circle tests executed, in the same conditions of speed and rudder angle. Empirical data are taken from The Specialist Committee on Esso Osaka, Final Report and Recommendations to the 23rd ITTC (The Specialist Committee on Esso Osaka, 2002).

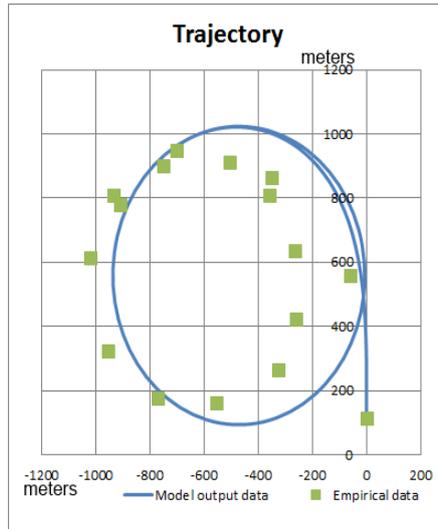
Figure 2: Turning trajectory in port side circle test, speed 8kn, rudder angle 35°.

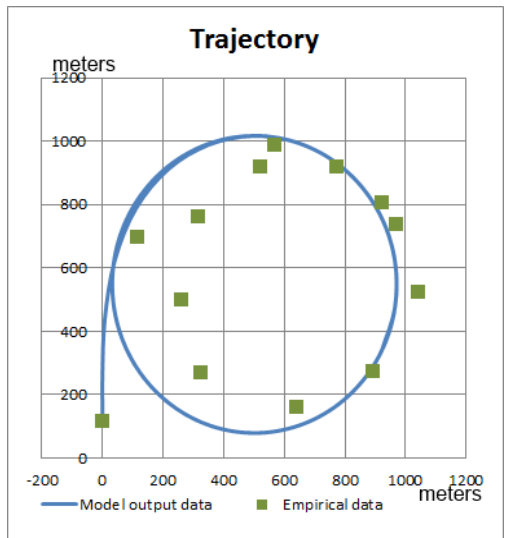
Figure 3: Turning trajectory in starboard side circle test, speed 10kn, rudder angle 35°

Figure 4 and 5 show two zigzag tests. They reflect the empirical data that can be seen on The Specialist Committee on Esso Osaka, Final Report and Recommendations to the 23$^{rd}$ ITTC.

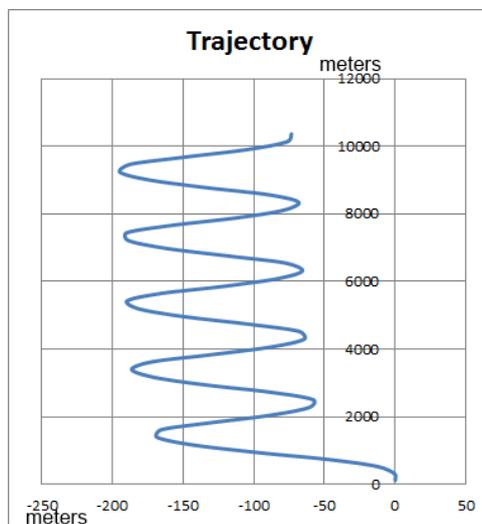
Figure 4: Turning trajectory in port side 20/20 zigzag test, speed 5kn

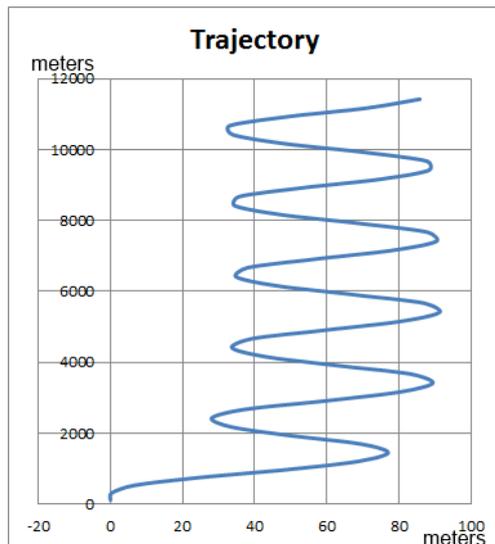
Figure 5: Turning trajectory in starboard side 10/10 zigzag test, speed 5kn

## 3. 3D GEOMETRIC MODELS AND VIRTUAL ENVIRONMENTS

The equations described in the previous section rule the 6 DOF motion of the ship at sea; such equations are only a part of the simulation based system prototype proposed in this article; simulators for training are characterized by 3d Virtual Environments able to provide the user with the sensation to be in the real system. Among others, currently the system includes the virtual environments of the port of Salerno (Italy). The port of Salerno, thanks to its central position in the Mediterranean Sea, has a crucial role in the national and international maritime trade. The port area includes the following quays:

- Quay of Ponente , length 563m, dockings n. 22-24;
- Rosso quay, length 226m, dockings n. 20-21;
- Trapezio quay, length 890m, dockings n. 13-19;
- Ligea quay, length 250m, dockings n. 11-12;
- 3 Gennaio quay, length 446m, dockings n. 7-10;

Outside the commercial area, on the east side of the port is located the Manfredi quay (length 380m, dockings n. 1-3) where a Marine Station devoted to cruise ships is being built. An entrance channel (280m large and 13m deep) with a 550 m diameter and 12 m deep evolution area characterizes the port. Moreover the port is served by 4 tugboats (working 24/24 h), 5 expert pilots with 2 equipped pilot-boats and 10 mooring operators with 2 equipped motorboats.

Vessels can pass across the entrance channel only one-by-one: if a boat is entering the port and another one is exiting, the incoming boat will wait until the other has safely completed its manoeuvres. In addition, due to Port Authority restrictions on side thrusters' use and wind conditions (that sometime can reach 40 kn), most of the large ships are obliged to ask the support of (at least) one tugboat during their manoeuvres. The figure 6 shows a panoramic view of the Port of Salerno.

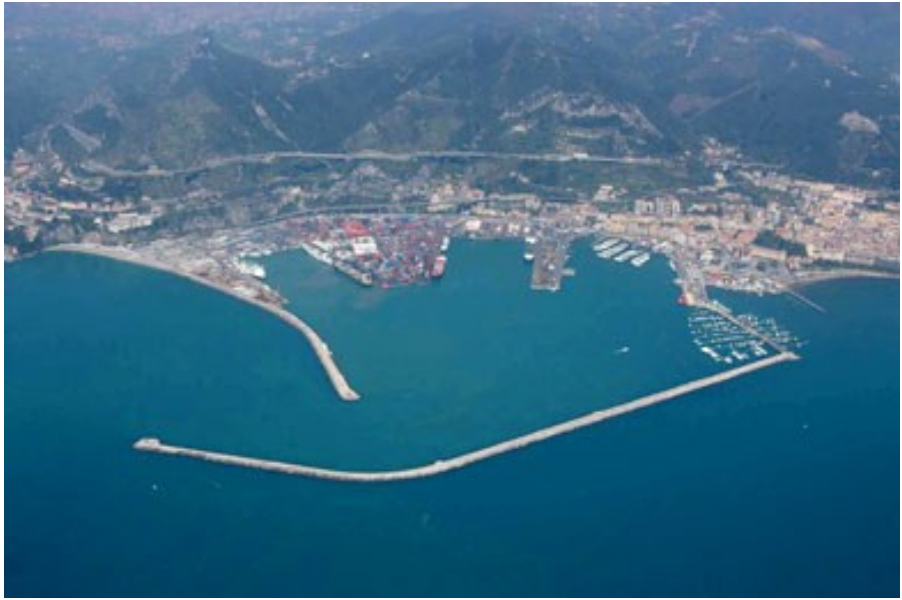
Figure 6 - Panoramic view of Salerno port

Control and support of ship navigation in the second last/ last mile of navigation are very complex activities. The high number of ships and vessels that usually enter to and exit from a marine port, the adverse visibility and weather conditions that may occur, the type of ship (i.e. container carriers, passengers ships, etc.), the docking times, the loading/unloading times, are examples of factors that slow down the traffic. These factors also increase the risk of collisions during the navigation in the harbour area. In this scenario, the proposed simulators aim at providing a concrete support by leveraging the role of pilots and port traffic controllers training. As a matter of facts, manoeuvring a ship while it is exiting from or entering into a port area is quite difficult and may cause terrible losses. To this end, it is crucial that ship pilots as well as port traffic controllers are well trained and fully aware of the consequences of their actions. Since simulation provides a safe training environment where trainees can gain experience and explore possibilities, simulation-based training has great potentials in this application domain. However, it is worth saying that the effectiveness of simulation based training increases as the trainee's feeling of realism increases. Hence, a great attention has been paid on the 3D geometric models and virtual environments that recreate the port of Salerno. The simulation software Creator and Vega Prime by Presagis have been used for creating highly optimized 3D geometric models and high-fidelity 3D real-time simulations respectively. As a result the simulator gives the possibility to steer a container carrier in the last mile of navigation, as well as offshore, setting the rudder angle, the propeller rate and a number of other parameters dynamically. Figures 7 shows the 3D geometric model of the port of Salerno, while figures 8 and 9 show two different views of the containership within the virtual environment of the port of Salerno.

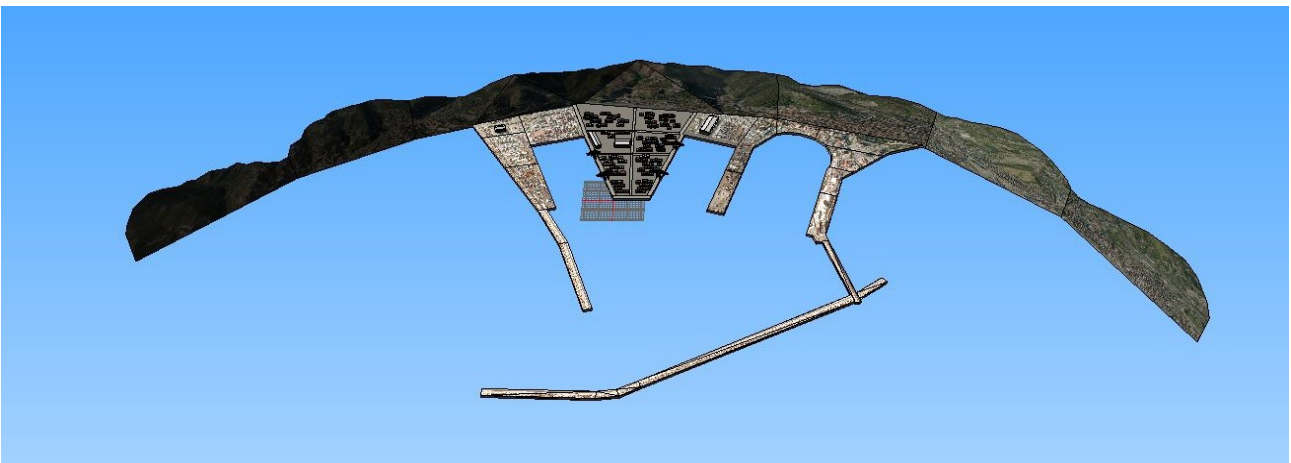
Figure 7 – 3D geometric model of the Port of Salerno

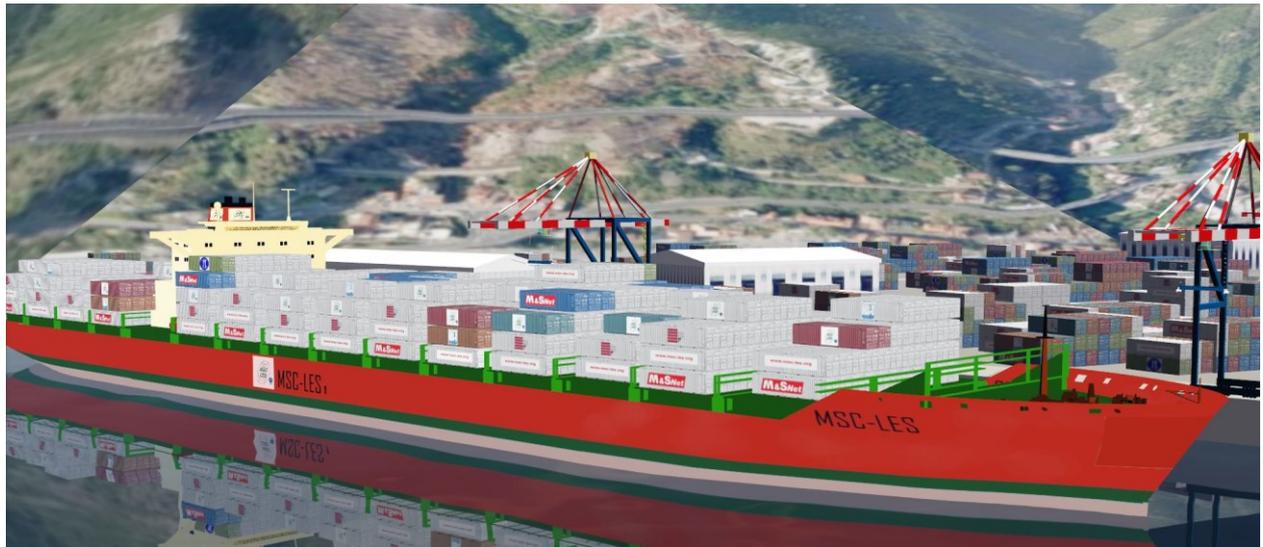
Figure 8: View on the Containership in the Virtual Environment of the Port of Salerno

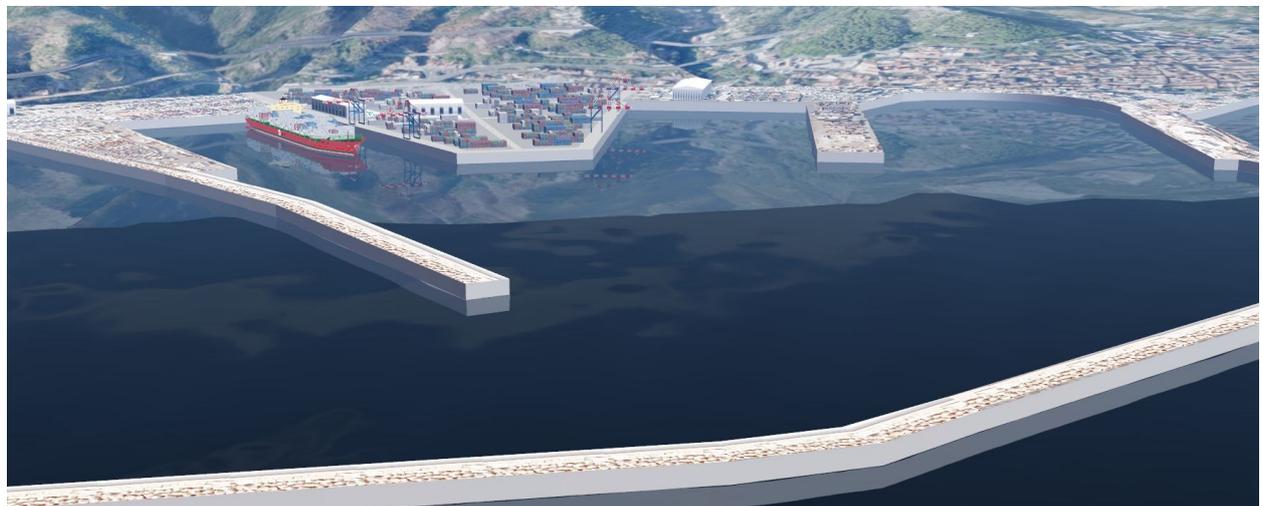
Figure 9: Panoramic view of the Virtual Environment of the Port of Salerno

## 4. THE SHIP BRIDGE REPLICA DESIGN AND HARDWARE INTEGRATION

As part of this project, a ship bridge replica has been designed and recreated including all the on board instrumentations used for ship navigation. The ship bridge replica, that is part of the system, has been designed to be a faithful representation of a real ship bridge. To this end, at the early stage of the design process some real ships were surveyed. In particular, much time was spent on board of two ships: a platform supply vessel that was standing in the port of Crotone (Italy) and a container carrier that was entered into the port of Salerno. These visits were really useful to have a faithful representation of the ship bridge in terms of geometry and technical equipment for ship manoeuvring and navigation.

### 4.1 The Ship Bridge Replica Design

As a result, to the ship bridge replica was designed with a T-shape geometry as illustrated in figure 10 where the 3D CAD model is depicted. Moreover, the overall dimensions, as reported in figures 11-13, have been set in order to easily integrate the hardware needed to control the motion of the ship at sea as well as to host the navigation instrumentations. Moreover, such dimensions have been also defined taking into account the need to comply with ergonomic requirements. Indeed, the trainee while standing in front of the instruments panel must be able to use and access all the control devices comfortably, i.e. the operator must be able to reach easily all the instrumentations available and must be able to look at the Virtual Environment from a suitable distance. As for the material, wood has resulted in a good choice with fairly low costs and good stability. It is worth mentioning that the overall bridge replica design has been carried out in-house while the manufacturing process has been carried out by a woodworker that is a specialist in ships and ship interiors building.

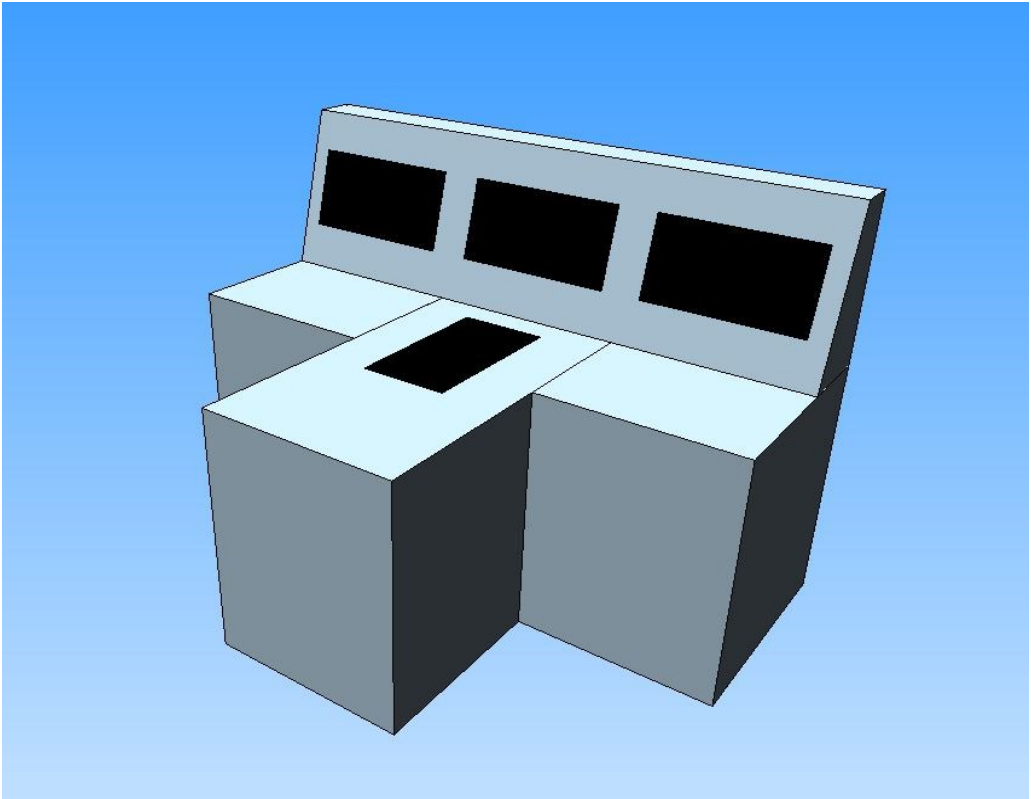
Figure 10: Prospective view of the ship bridge replica CAD

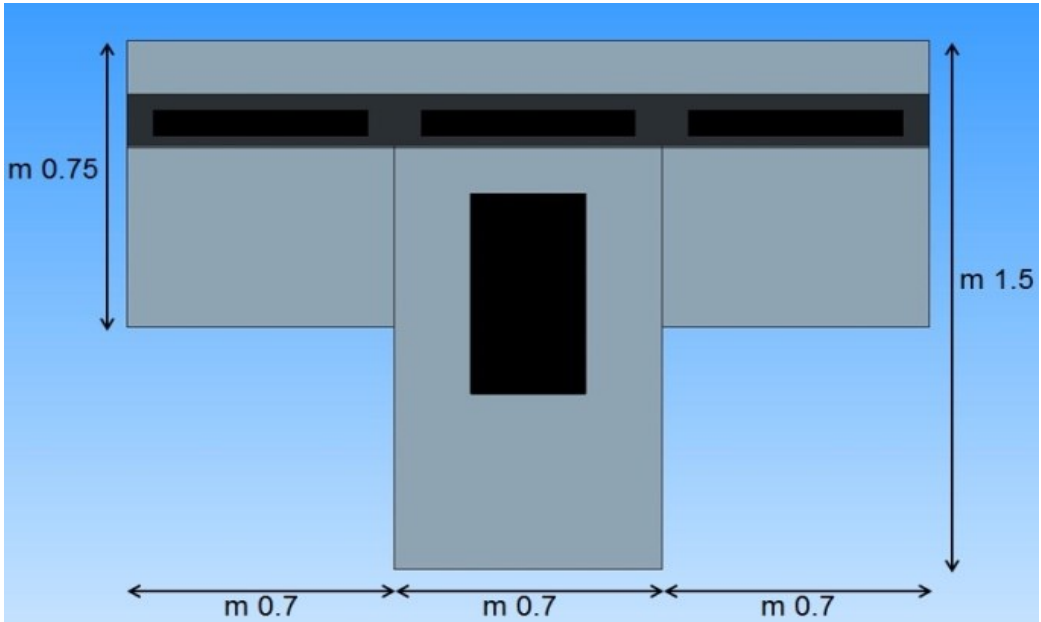
Figure 11: Top view and dimensions of the ship bridge replica CAD.

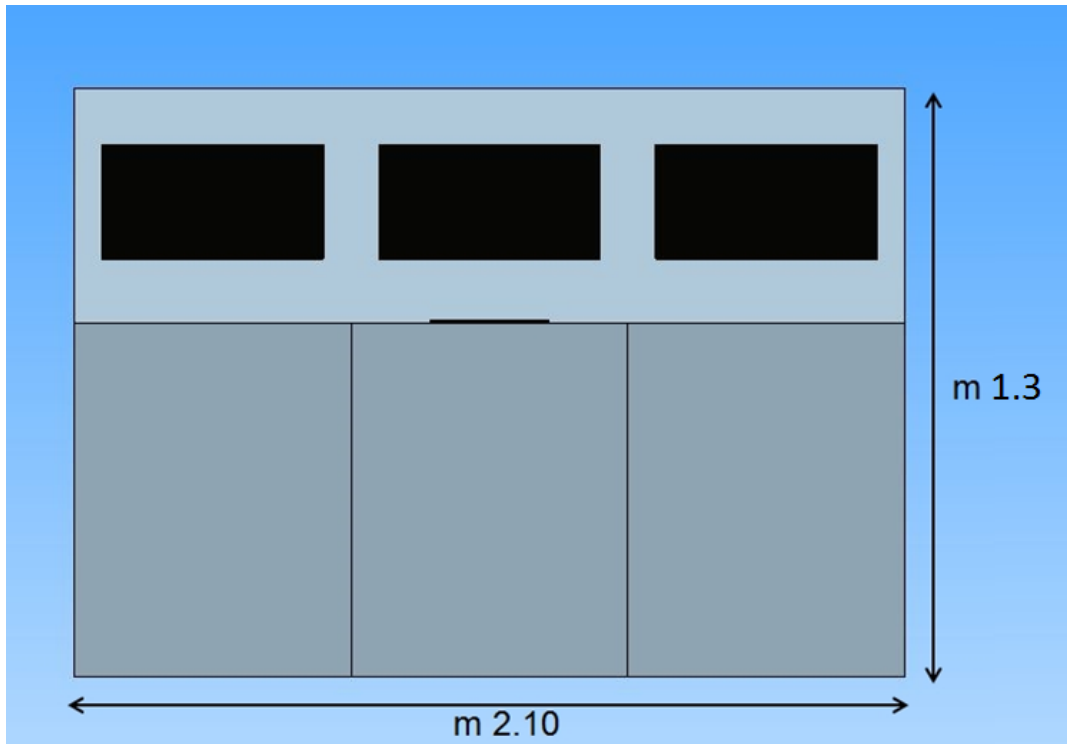
Figure 12: Front view of the ship bridge replica CAD

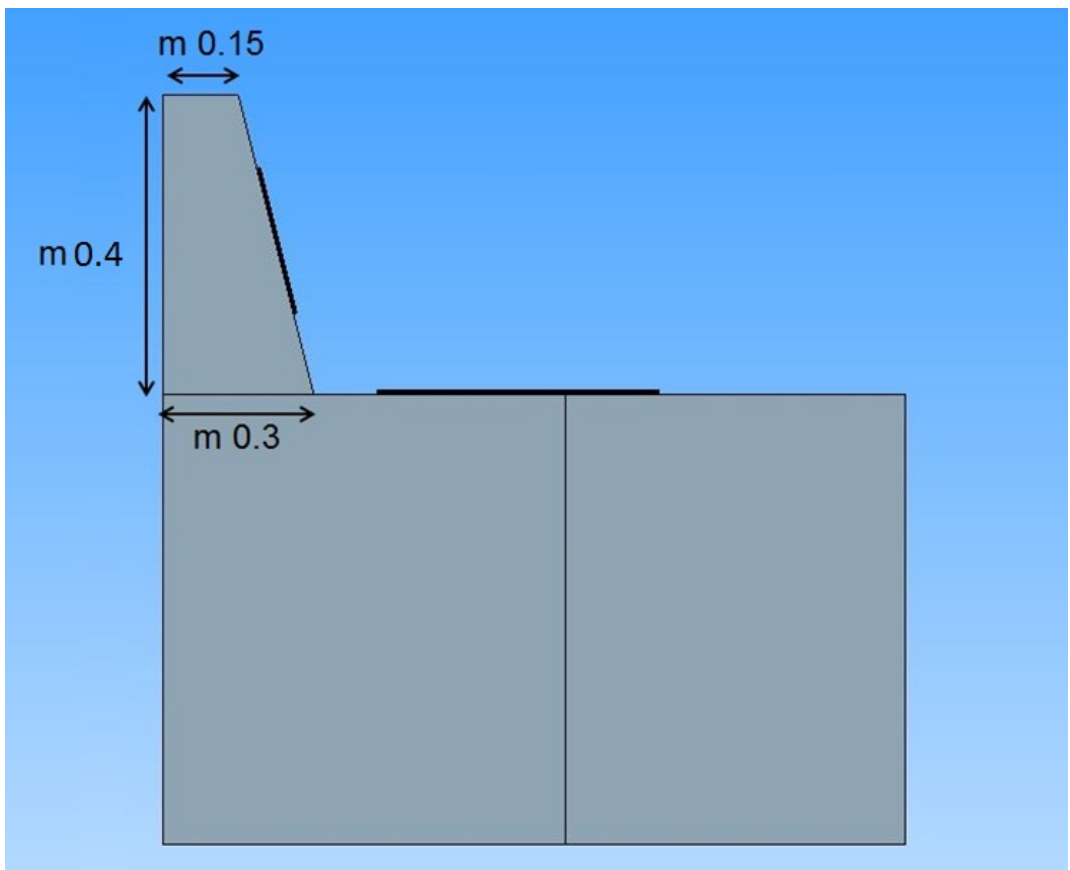
Figure 13: Left view of ship bridge replica CAD

The ship bridge replica can host up to three PCs that can be accessed by three doors placed in the rear part (as explained later on in the article, currently thanks to a strong modelling effort and computational and graphic load optimization the entire simulation including the ship motion at sea in the virtual environments, the AIS simulator, the

Radar simulator and the Conning display are able to run on a single commercial desktop computer). Moreover in figures 10 and 11 four black rectangles are depicted; three of them placed in the upper part are the slots for the AIS simulator, the Conning information display, the Radar simulator, while the last one hosts a touch screen that allows interacting real-time with the simulation environment and can be used to change the simulator viewpoints, edit the camera of each view, drop/weigh anchor and control the navigation instrumentations. Moreover, the ship bridge replica is also integrated with mechanic and electrical devices that recreate the manoeuvring controls: it includes a two-lever system for engine RPM and propeller pitch controls (in order to simulate ships with both fixed and variable pitch propellers), a double joystick for side thrusters control and the wheel for rudder control. It is worth mentioning that all the aforementioned devices have been bought form shops that are specialized in manufacturing and selling spare parts for ships and they have been opportunely adapted and integrated as part of the simulation system. The hardware integration is a very important feature of the training system since it allows the trainee to be acquainted with the same equipment available on real ships and, as a consequence, to benefit from a greater level of immersion during a training session. As a matter of facts, such devices are fully integrated with the simulation environment; in other words the user can have a real time response that is displayed on the visualization system based on the input he/she provides while using ship controls for manoeuvring. The interfaces between on board instrumentations and the PC that drives the simulation have been designed and implemented at the MSC-LES lab by using the Arduino technology. Arduino is also used to control multiple LED systems and three small-size displays that are used to show the rudder angle, the main engine RPM, the main propeller pitch and the side thrusters power. Figures 14-16 show the ship bridge controls and display systems in their current configuration.

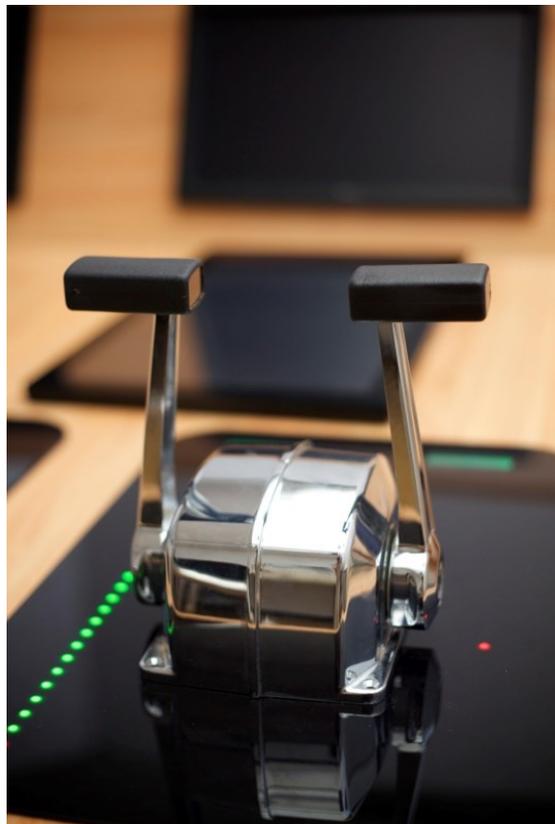

Figure 14: Ship bridge detail: double lever system and multiple LEDs displays

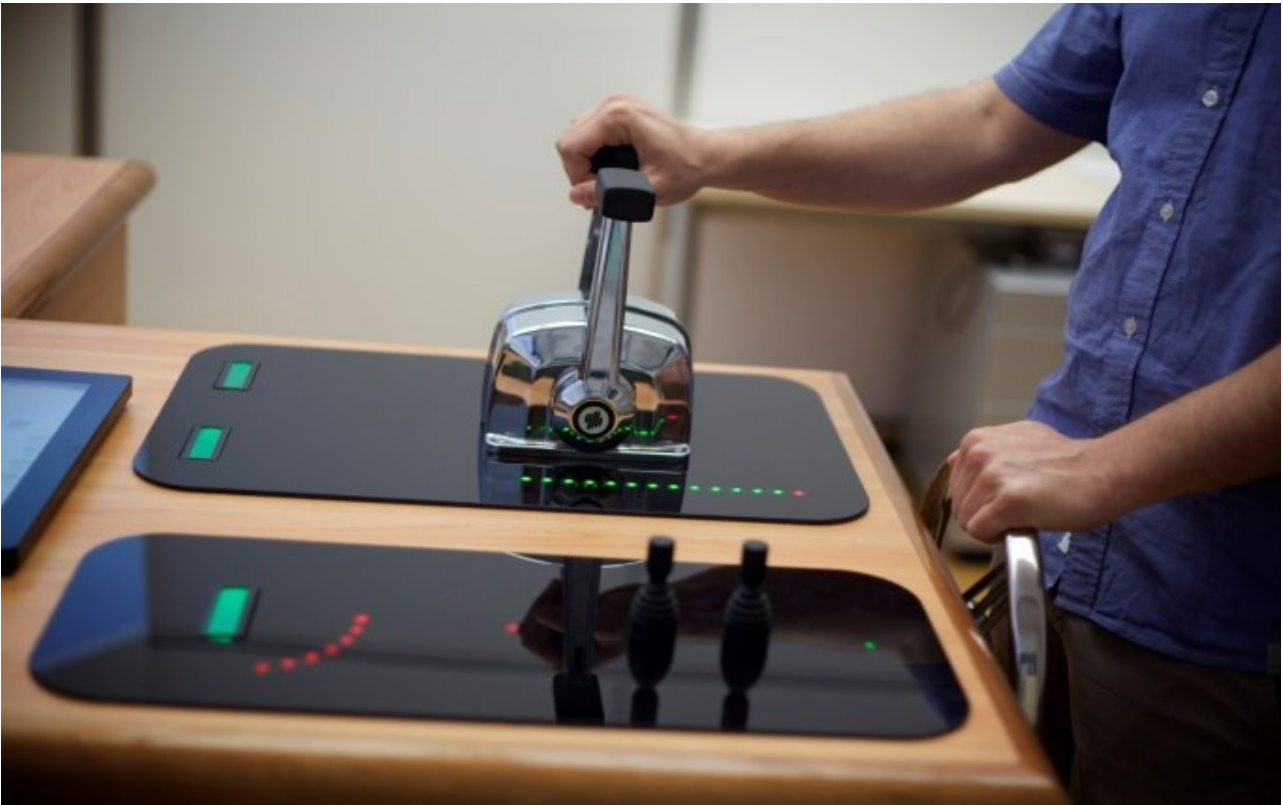

Figure 15: Ship bridge detail: double lever system, joysticks for side thruster control, wheels for rudder control, multiple LEDs displays and small-size displays

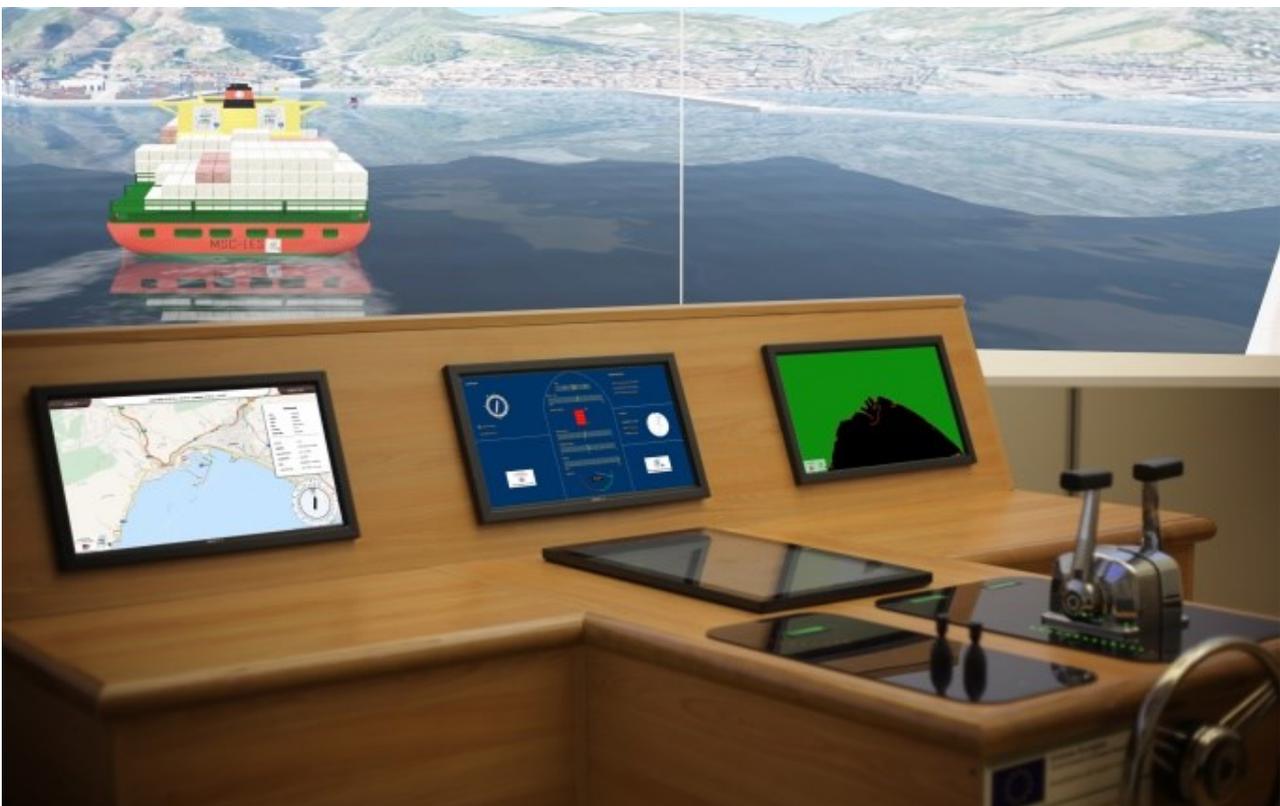

Figure 16: Ship bridge detail: navigation instruments and visualization area including AIS simulator, Conning display simulator, Radar simulator

## 4.2 The Visualization System for the Virtual Environments, the Sound System and PCs requirements

The visualization system shows the trainee how the scenario evolves over time as a consequence of the actions he/she undertakes. For the sake of increasing the user's level of immersion, the visualization system includes three large white screens and three projectors (partially visible in figure 16). Each screen has 2 x 1.5 m dimensions without external frame; one of the screens is place behind the ship bridge just in front of the trainee's position while the other screens are placed on the left and on the right side respectively. Moreover, the screens are juxtaposed to ensure a seamless visualization and enhance the trainee's feeling of immersion in the simulation (figure 16 partially shows how screens are positioned with respect to the ship bridge and the trainee position). In order to increase the flexibility and reusability of the simulation system, dedicated iron frames were designed and manufactured to hold up both screens and projectors. This is resulted in a very flexible and portable solution, since the whole visualization system including screens and projectors, can be easily moved from its current location to another as well as adapted to different configurations in terms of dimensions of the projected area and position (to be used for different virtual simulation based systems).

Both for the screens and the projectors the frames are made of iron in order to ensure solidity and stability while keeping the costs as low as possible. The CAD model of the screens supports and their geometry is depicted in figure 16. Furthermore, as shown in figure 17, the iron frame is 2.5 m in height and it lies on two feet (each foot is made up of 0.5 m pieces of square tube).

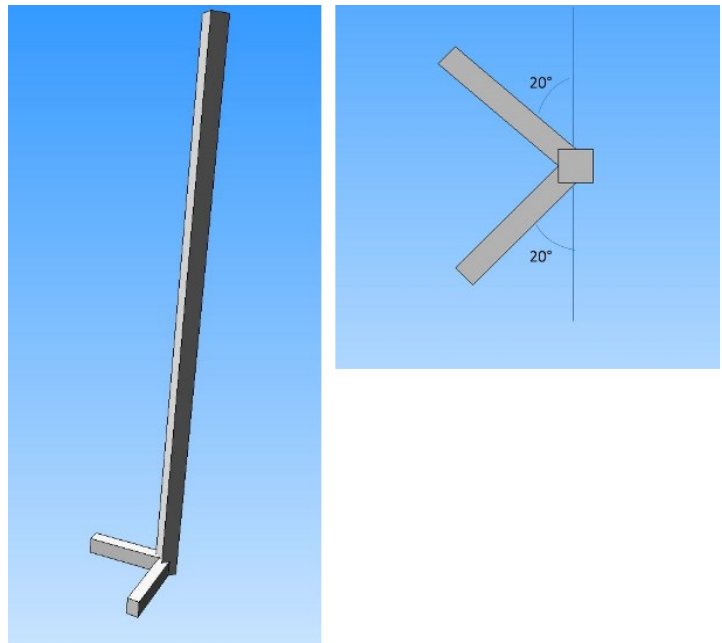

Figure 17: Prospective and top view of the foot for screen support CAD

For the projectors (see figure 18) the support structure is telescopic for both the width and the height in order to ensure different configurations in terms of dimensions of the projected area and position. The structure is composed of two vertical beams and one crossbeam. The vertical beams are hold up by two square tubes of 0.60 m length. Each vertical beam is actually composed by two square tubes, one with a side section of 5 cm and the other with a 4 cm section in order to create a telescopic system. The two beams can scroll one inside the other and this allows the user to choose the right length of the vertical beam. The same working principle has been adopted for the crossbeam, but, in this case, it is composed by two square tubes with a 5 cm side section and a 4 cm side section. Thanks to this double telescopic solutions (on the height and on the width), it is possible to modify the height of the structure between 2.2 meters and 3.2 meters and the width between 3.5 meters and 5 meters. In addition, all the video and electrical cables can be easily passed within the square tubes. Figure 18 shows the prospective view of the CAD of the entire structure. The figure 19 shows a view of the projectors installed on the structure.

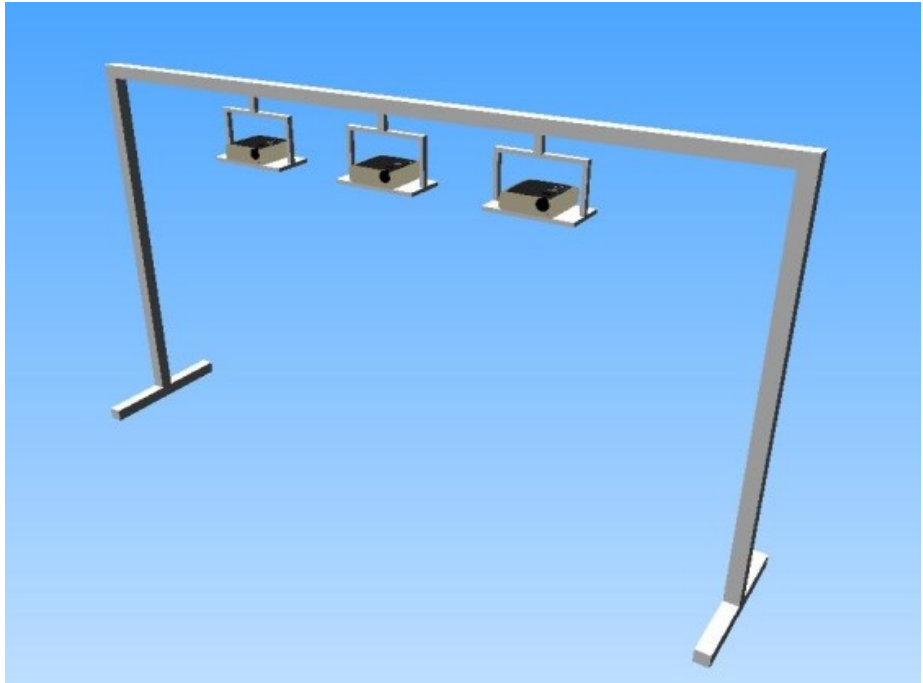

Figure 18: Prospective and top view of the CAD of the structure for projectors

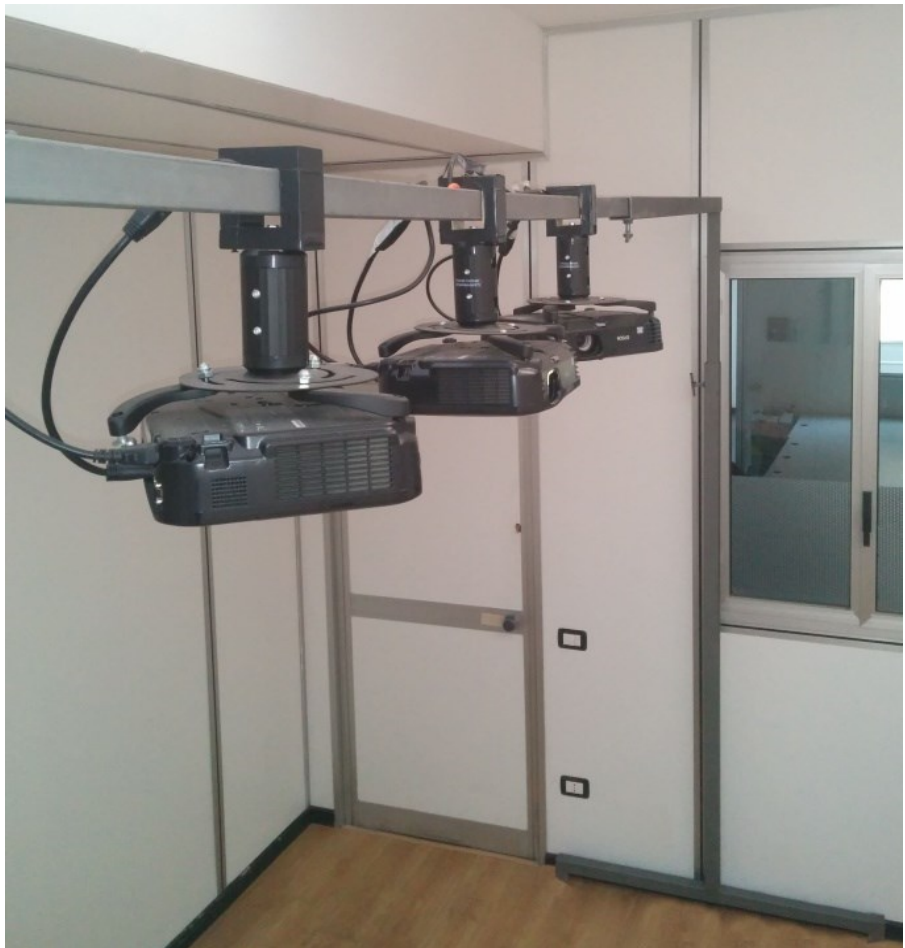

Figure 19: View of the projectors installed on the structure

Finally, a 5.1 surround sound system has been added able to reproduce sounds usually audible in the ship cockpit. The 5.1 surround system contributes in increasing the feeling of being inside a real ship.

As far as the PCs requirements are concerned, one of the main goals that was defined at the early stage of the design process was that the architecture of the simulation system had to be light enough and flexible in order to allow the simulation system working with any well-equipped commercial desktop PC. The main requirements of the PC that is currently used inside the bridge ship replica are given below:
- Processor: Intel Core i7 2600k 3.40 Ghz
- RAM: 16 Gb
- GPU: 2 GeForce GTX 680 connected with SLI technology.

The use of two GPU depends on the need to manage 4 monitors and 3 projectors at the same time with one PC.

**5. THE CONTROL TOWER SIMULATOR**

As already mentioned the simulation system is devoted to cooperative training of ship pilot and port traffic controllers in the last mile of navigation. This section briefly presents the control tower simulator. Such simulator is connected to the ship simulator through the standard for distributed simulation IEEE High Level Architecture (HLA), therefore the two simulators (ship simulator and control tower simulator) are able to interoperate each other, sharing the same virtual environments and allowing cooperative training of ship pilots and port traffic controllers.

The Control Tower Simulator is equipped with a multi-display system, which displays the virtual environments for monitoring and controlling all the operations in the port area. The port traffic controller may act on the console to change the view (multiple panoramic views are available from different cameras placed in different positions within the virtual environments), to activate communications with the ship pilot, to display vessels traffic on 2D maps (similar to the AIS system) together with the navigation information of each vessel (e.g. current speed, current route, position, etc.), to activate the radar view.

In addition, the port traffic controller may activate a particular view: the so-called teleportation view. The teleportation view is the same view of the virtual environment perceived by the ship pilot (the view from the ship bridge). Therefore, by activating this view the port traffic controller can be teleported on board each vessel that is currently manoeuvring within the port area to observe all the operations from the same view and perspective of the ship pilot. This tremendously increases the effectiveness of the training for both the port traffic controller and the ship pilot because they can exchange information and look at the procedures performed (and related errors) from the same point of view.

Finally the port traffic controller is able to monitor information on weather and sea conditions.

The figure 20 shows the control tower simulator workstation with a panoramic view on the Port of Salerno; the control tower simulator workstation is equipped with three monitors that can be used for visualizing the 3D virtual environments as well as to display 2D maps, radar and all the other simulator functionalities.

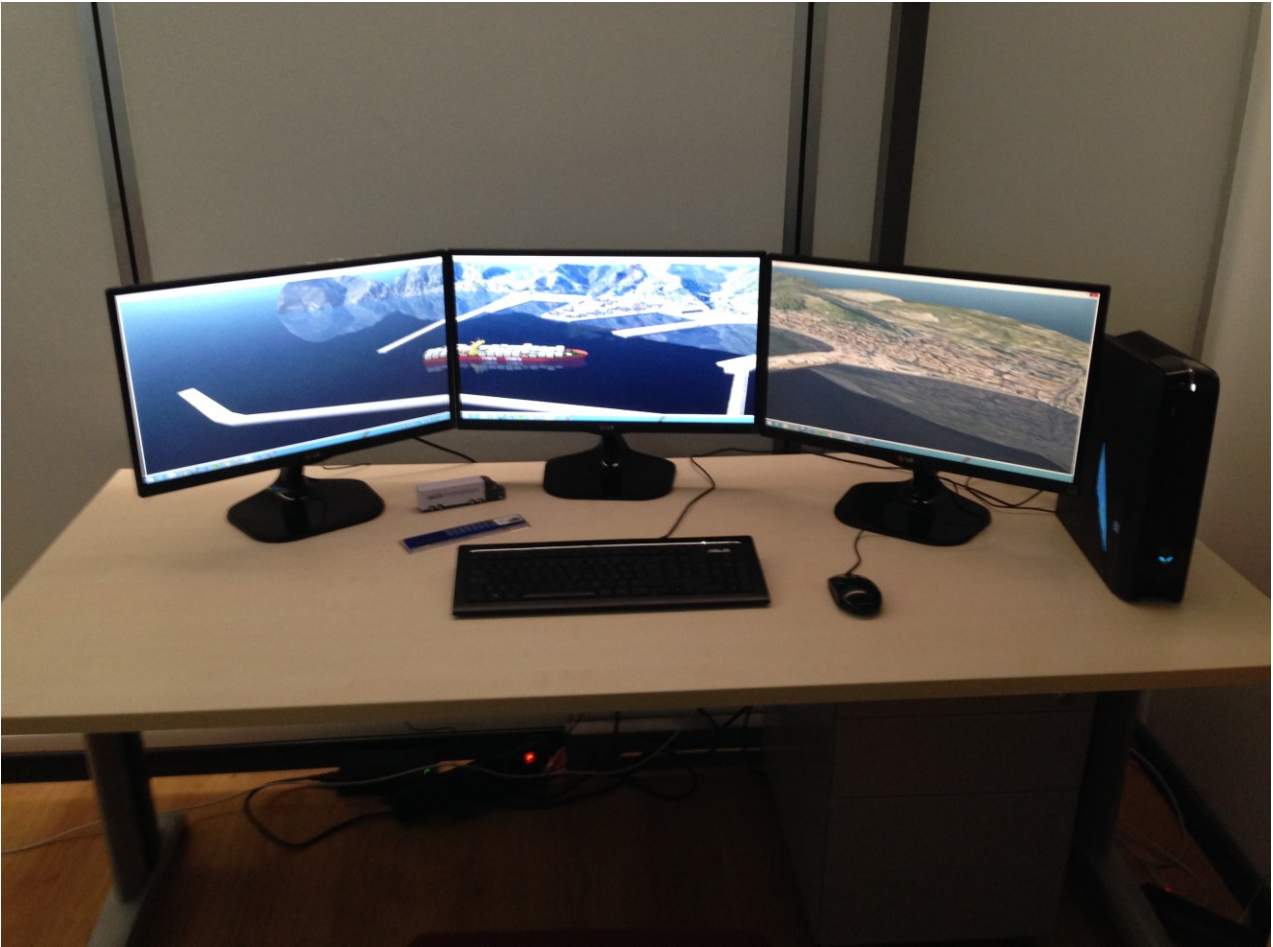
Figure 20 – The Control Tower Simulator

## 5. THE OVERALL PROTOTYPE SYSTEM

The simulation based system presented in this article has been conceived in order to provide the users with a large spectrum of operative scenarios; in fact, the instructor can set weather and marine conditions before the training session starts or even during the execution. Data such us main engine RPM, side thrusters utilization levels, rudder positions, wind intensity and directions, compass, 2D maps of the port areas including the other ships (Automatic Identification System – AIS) are always available during the simulation. Performance measures such as mission time (average and standard deviation values over multiple training sessions), collisions, and wrong manoeuvres are recorded and are available at the end of the simulation. Furthermore, different viewpoints are available during the simulation: inside the bridge, outside the ship (therefore it is possible to see the whole ship from different points of view). In addition, the accuracy and quality of the simulator has been guaranteed by Verification, Validation and Accreditation (VV&A) processes that have been carried out during the entire development period. As mentioned before, the ship dynamic has been verified and an ad-hoc tool has been developed to this purpose. As for the computer simulation model a verification has been carried out by using the debugging technique to ensure high levels of accuracy and quality.

The figure 21 shows the ship bridge simulator in use while the containership is entering the port of Salerno (on the bridge ship simulator the camera is showing a rear view of the ship; multiple view are allowed as shown in figure 22 where the bridge view is displayed).

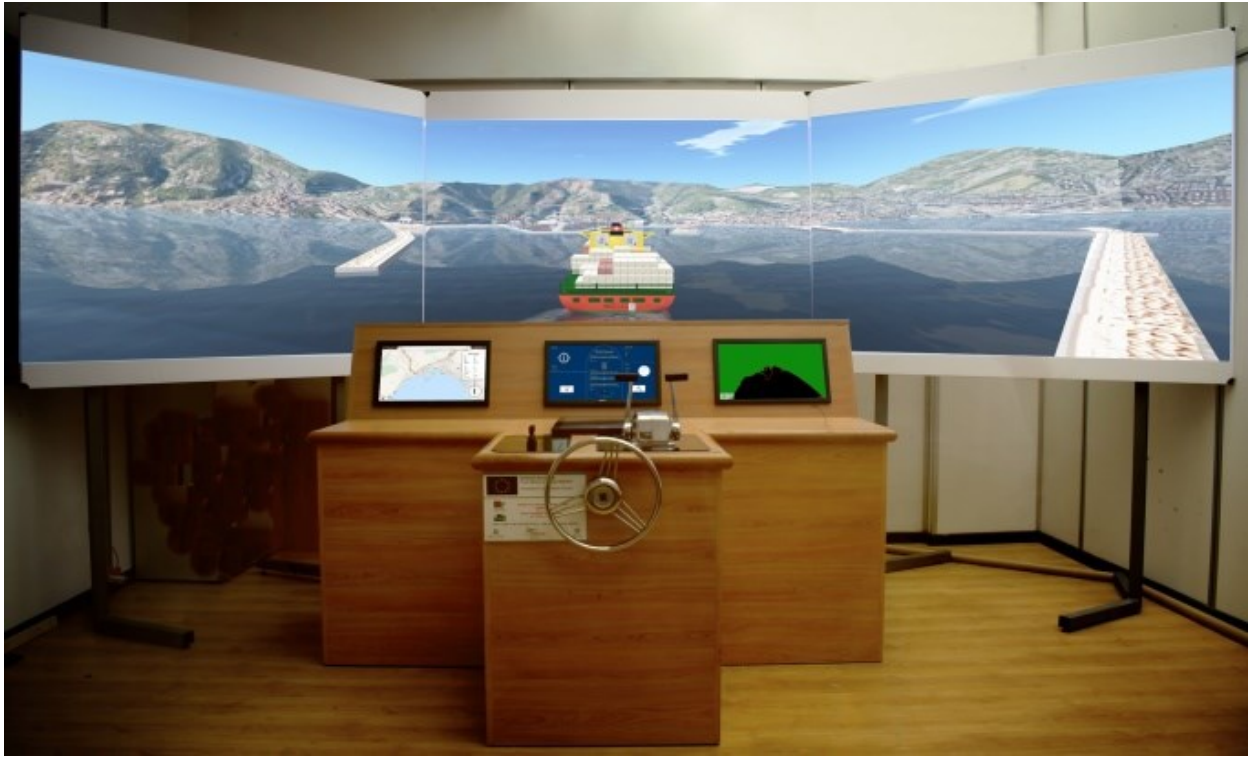
Figure 21: The ship bridge replica and the visualization area

The figure 22 shows the full system prototype in use, including two simulators (the ship bridge simulator on the left and the control tower simulator on the right) working together (interoperability mode) and sharing the same virtual environments thanks to the use of the IEEE 1516 HLA standard for distributed simulation. In figure 22, a view from the bridge of ship is shown (the view that normally perceived by the pilot while performing manoeuvres within the harbour area).

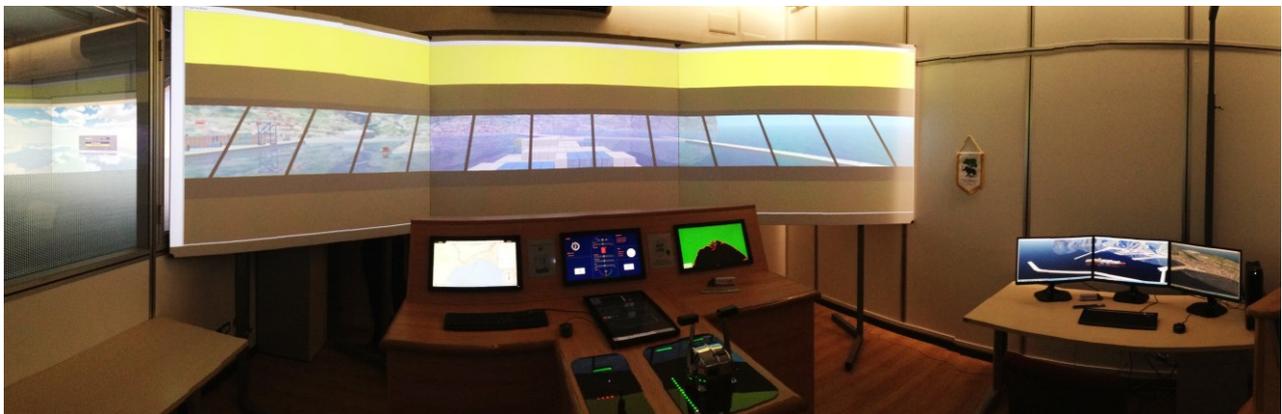
Figure 22 – The ship bridge simulator and the control tower simulator working together for cooperative training of ship pilots and port traffic controller (Interoperability use mode and distributed simulation through High Level Architecture)

## 5. CONCLUSIONS
This article presents some of the results of the HABITAT research project carried out at MSC-LES lab of University of Calabria. The main goal of this research project is to develop and prototype a fully operative simulation based training system for cooperative training of ship pilots, port traffic controllers and tugboat pilots. In this paper, the design and development of the ship bridge simulator and control tower simulator are presented. As far as the bridge ship simulator is concerned, the article faces and solves different problems, namely: the motion of the ship at sea, the design and development of the 3D geometric models and virtual environments, the hardware design and integration. To this end, the equations for a 6 DOF model for the ship motion at sea are presented, discussed and validated. The 3D geometric models and virtual environments that recreate, with high level of detail, a real port (the Port of Salerno) are presented. Finally the article presents the design of the ship bridge replica, the integration of the hardware needed to control the ship motion at sea and the additional instruments to support the navigation (AIS simulator, Radar simulator and conning display system simulator).

As far as the control tower simulation is concerned, the article briefly presents the main functionalities of the simulator and its integration through the IEEE 1516 HLA standard for distributed and interoperable simulation. Even though several ship simulators are already on the market, few of them offer the possibility of setting up combined and integrated training sessions involving both ship pilots and port traffic controllers. In fact, the last mile of navigation has not been subject of particular scientific interest until the recent disaster in the port of Genoa (the Jolly Nero ship accident). Considering these aspects the proposed simulation based training system takes advantages of the benefits provided by Modeling & Simulation and allows:

- improving the trainees skills in performing complex manoeuvres within the port area
- learning about the procedures adopted in a certain port (it is important in this sense to underline the possibility to set-up different virtual environments recreating additional marine ports);
- improving synergy and communication between ship pilots and port traffic controllers;
- defining new policies and new procedures and testing their effectiveness.

Additional research activities are still ongoing and are focused on the development of the tugboat simulator and its integration in the current federation of simulators.

## ACKNOWLEDGMENTS

The research work described in this paper is part of the research project, HABITAT PON01_1936, co-financed by the Italian Ministry of Research and Education (MIUR) under the program "PON Ricerca e Comptetitività".